\documentclass[conference, compsoc, letterpaper]{IEEEtran}

\usepackage[english]{babel}
\usepackage{amsmath,amsfonts}
\usepackage{algorithmic}
\usepackage{graphicx}
\usepackage{textcomp}
\usepackage{csquotes} % for proper quotations
\usepackage{siunitx} % for proper units
\usepackage{xspace}
\usepackage{array}
\usepackage{caption}
\usepackage{epsfig}
\usepackage{endnotes}
\usepackage{graphicx}
\usepackage{booktabs}
\usepackage{multirow}

\usepackage{url}

\usepackage{hyperref}
\usepackage{xcolor}
\usepackage{subfigure}
\usepackage[ruled]{algorithm2e}
\usepackage{pifont} % for number in algorithm
\usepackage[capitalize,noabbrev]{cleveref}
\usepackage{adjustbox}

\usepackage[para,flushleft]{threeparttable}
\usepackage[acronym,nonumberlist]{glossaries}
\makeglossaries
\usepackage{enumitem}
\setitemize{wide}

\hypersetup{
    colorlinks,
    linkcolor={red!50!black},
    citecolor={blue!50!black},
    urlcolor={blue!80!black}
}

\crefname{section}{\S}{Sections}
\crefname{page}{page}{pages}
\crefname{paragraph}{\S}{Sections}
\Crefname{section}{Section}{Sections}
\crefformat{section}{\S#2#1#3}
\Crefformat{section}{Section~#2#1#3}

\crefname{equation}{}{}

\usepackage[symbol]{footmisc}

\newcommand{\name}{SAGE\xspace}
\newcommand{\nameacr}{\name: Software-based Attestation for GPU Execution}

\newcommand{\fdelim}{\ensuremath{\;\|\;}\xspace}

\newcommand{\mess}[3]{{\mathit{#1} \rightarrow \mathit{#2}} : \mathit{#3}}

\newcommand{\R}{\mathcal{R}}

\newcommand{\etal}{et~al.\@\xspace}

\usepackage{pifont}% http://ctan.org/pkg/pifont
\newcommand{\cmark}{\ding{51}}%
\newcommand{\xmark}{\ding{55}}%
\newcommand{\nvidia}{NVIDIA\xspace}

\DeclareSIUnit{\kB}{KB}
\DeclareSIUnit{\KB}{\kB}
\DeclareSIUnit{\MB}{\mega\byte}
\DeclareSIUnit{\mbps}{Mbps}
\DeclareSIUnit{\gbps}{Gbps}
\DeclareSIUnit{\tbps}{Tbps}
\DeclareSIUnit{\mpps}{Mpps}
\DeclareSIUnit{\kBps}{kB/s}
\DeclareSIUnit{\b}{b}
\usepackage{mathtools}

\usepackage{titlesec}
\titlespacing*{\section}{0pt}{2ex}{1.4ex}
\titlespacing*{\subsection}{0pt}{1.0ex}{0.8ex}
\titlespacing*{\subsubsection}{0pt}{0.8ex}{0.6ex}

\renewcommand\paragraph[1]{\smallskip\textit{#1.}}
\newskip\smallskipamount \smallskipamount=1pt plus 1pt minus 1pt

\usepackage{enumitem}
\setlist[itemize]{itemsep=0.1pt}
\setlist[itemize]{nosep}
\setlist[enumerate]{itemsep=0.1pt}
\setlist[enumerate]{nosep}
\setlist[enumerate]{leftmargin=*}

\makeatletter % changes the catcode of @ to 11
\newcommand{\linebreakand}{%
  \end{@IEEEauthorhalign}
  \hfill\mbox{}\par
  \mbox{}\hfill\begin{@IEEEauthorhalign}
}
\makeatother % changes the catcode of @ back to 12

\begin{document}

\title{\nameacr}

\author{\IEEEauthorblockN{Andrei Ivanov\IEEEauthorrefmark{1}\IEEEauthorrefmark{2}, Benjamin Rothenberger\IEEEauthorrefmark{1}\IEEEauthorrefmark{2}, Arnaud Dethise\IEEEauthorrefmark{3}, Marco Canini\IEEEauthorrefmark{3}, Torsten Hoefler\IEEEauthorrefmark{2}, Adrian Perrig\IEEEauthorrefmark{2}}
\and

\linebreakand

\IEEEauthorblockA{\IEEEauthorrefmark{2}\textit{ETH Z\"urich} \\
firstname.lastname@inf.ethz.ch}
\and
\IEEEauthorblockA{\IEEEauthorrefmark{3}\textit{KAUST} \\
\{firstname.lastname, marco\}@kaust.edu.sa}
}

\date{}

% \keywords{GPU, attestation, software}

\maketitle

\thispagestyle{plain}
\pagestyle{plain}

\footnotetext{*Equal contribution}

\begin{abstract}
With the application of machine learning to security-critical and sensitive domains, there is a growing need for integrity and privacy in computation using accelerators, such as GPUs.
% Machine learning often requires the use of accelerators such as GPUs to handle the immense computational workload required by complex applications.
Unfortunately, the support for trusted execution on GPUs is currently
very limited -- trusted execution on accelerators is particularly
challenging since the attestation mechanism should not reduce performance.
% hinder their high performance requirements.

Although hardware support for trusted execution on GPUs is
emerging, we study purely software-based approaches for
trusted GPU execution. A software-only approach offers distinct
advantages: (1)~complement hardware-based approaches, enhancing
security especially when vulnerabilities in the hardware
implementation degrade security, (2)~operate on GPUs without hardware
support for trusted execution, and (3)~achieve security without
reliance on secrets embedded in the hardware, which can be extracted
as history has shown.

In this work, we present \name, a software-based attestation mechanism
for GPU execution. SAGE enables secure code execution on NVIDIA GPUs
of the Ampere architecture (A100), providing
properties of code integrity and secrecy, computation integrity, as
well as data integrity and secrecy -- all in the presence of malicious
code running on the GPU and CPU. Our evaluation demonstrates that
\name is already practical today for executing code in a trustworthy
way on GPUs without specific hardware support.

\end{abstract}


\section{Introduction}

Fueled by recent trends such as machine learning and the declining
yields from Moore's Law, the use of accelerators
% such as GPUs, 
to process the vast volumes of data is becoming indispensable. In fact,
it is expected that the
majority of compute cycles in public clouds will be executed on
accelerators~\cite{acc-market}.

%% This trend towards the use of accelerators to offload compute tasks,
%% GPUs are used to perform increasingly complex work and have become
%% powerful execution environments with thousands of processing cores and
%% high memory bandwidth. GPUs are also becoming widely available with
%% their adoption into SoCs on mobile devices and large-scale deployments
%% in the cloud.

With the application of machine learning to security-critical or sensitive
domains such as healthcare or financial modeling, there is a growing need 
for a mechanism that maintains integrity and secrecy for both code and data despite the
computation being offloaded to the GPU.

With the wide-spread deployment of trusted execution environments
(TEEs), e.g., Intel SGX~\cite{costan2016intel} and ARM
TrustZone~\cite{arm-trustzone}, an important question is how
security-sensitive computation tasks can be accomplished on GPUs.
While first hardware-based TEEs on GPUs are starting to
emerge~\cite{volos2018graviton, jang2019heterogeneous,zhu2019enabling,nvidia-mig,nvidia-egx, olson2015border}, how can we execute code
securely on GPUs \textit{in current environments}? As we have
witnessed from the introduction of hardware-based TEEs on x86
platforms, it took over a decade until it became possible to fully and
widely utilize these mechanisms. 
At the same time, technology progress in this space is a moving target as new attacks (among other factors) force vendors to phase out one specific hardware-based technology in favor of more robust successors (such as with the case of the deprecation of Intel SGX~\cite{sgx-deprecation}). 
Given the importance of software executing on GPUs,
it is clear that we need to find approaches to speed up the long lag
time between deployment and wide-spread utilization.

A promising approach for bridging this gap is a software-only approach
to trusted execution. In the context of CPU-based execution, a rich
research field has contributed numerous
approaches~\cite{gligor2019establishing, seshadri2005pioneer, seshadri2004swatt, zhao2013redabls}. The basic approach of the prior
software-based or timing-based attestation approaches was to design a
verification function that would run on an untrusted system and
compute a checksum over itself -- where both the correctness of the
checksum and the time duration are measured by a trusted verifier. A
correct checksum value that is returned before a threshold point in
time, indicated to the verifier that the TEE was correctly set up and
that the correct code is now executing (code integrity and launch
point integrity). In combination with a system for control-flow
verification, control-flow integrity can also be achieved.

The challenge of such software-based TEE establishment approaches lies
in the creation of a verification function that will slow
down noticeably or produce an incorrect checksum, if an adversary
attempts to tamper with its execution.

The creation of a verification function for GPU environments poses
numerous research challenges, which may be the reason why it has so
far not been achieved, to the best of our knowledge. First and
foremost, achieving (1) code secrecy and integrity, and (2) data
secrecy and integrity, (3) in the presence of a malicious OS, (4)
malicious code on GPU, and (5) a malicious CPU-GPU interconnect is a
formidable challenge. Other challenges that we had to overcome include
the absence of a true random number generator on the GPU, the lack of documentation
from GPU vendors for a specific target architecture, no toolchain support to write
native GPU code, and achieving an optimal GPU utilization.

We design the \name system, which establishes a TEE on \nvidia GPUs of
the Ampere architecture (A100). \name utilizes an SGX
enclave running on the host to act as a local verifier, and to bootstrap the
software primitive to establish a dynamic root-of-trust (RoT) on the GPU. RoT
establishment ensures either that the state of an untrusted system contains all
and only content chosen by a trusted local verifier and the system code 
begins execution in that state, or that the verifier discovers the existence of
unaccounted content. \name also sets up a shared secret key
between the verifier and the GPU, which can be used to 
establish a secure channel to achieve
integrity and secrecy for code and data transferred. Our results
indicate that after a successful invocation of \name, the verifier obtains assurance that: (1) the
user kernel on the untrusted device is unmodified; (2) the user kernel is invoked
for execution on the untrusted device; and (3) the executable is executed
untampered, despite the potential presence of a malicious actor.

This paper presents the following contributions:
\begin{itemize}[nolistsep]
    \item We design a software-based attestation mechanism for GPU execution that enables secure code execution on \nvidia Ampere GPUs, providing code integrity and secrecy, computation integrity, as well as data integrity and secrecy.
    \item We implement an instruction generation framework that is capable to generate GPU microcode. This requires understanding the GPU architecture and the instruction format used in microcode, which we obtained using a instruction decoding framework.
    \item We implement a true random number generator (TRNG) for execution on GPU based on race-conditions.
    % \item We evaluate the performance of our VF design and show that the optimized checksum function is around 250\% faster than a regular implementation with highest level of optimizations enabled.
    % We also show that with \ben{xx} iterations of the checksum function, an adversary is unable to inject additional instructions without causing a detectable overhead.
    % \item We formally verify the key-establishment protocol between the verifier and the GPU using the Tamarin prover~\cite{tamarin}.
    \item Through a proof-of-concept implementation on the \nvidia A100 platform, we demonstrate the technical feasibility of the approach.
\end{itemize}

\section{Background: GPU Fundamentals}
\label{sec:background}
In the following, we describe the fundamentals of \nvidia GPUs and their
programming model (CUDA) to illustrate how compute tasks are offloaded and
executed on the GPU. We focus on mechanisms relevant to this work.

The GPU is connected via the PCI control engine to the host CPU and uses an
internal bus for communication between its core components. The core components
are the command processor, compute and DMA engines, and the memory system, consisting of a
memory controller, registers, on-chip and device memory.

\paragraph{Controlling the GPU}
Commands to the GPU are transmitted using a set of command queues known as
\emph{channels}. The GPU's command processor receives these commands and forwards
them to the corresponding engines.

\paragraph{Data transfer to the GPU}
GPU programming inevitably incurs data transfers between host and device memory.
This is handled using direct memory access (DMA). The copy engine is responsible
for handling DMA commands and their corresponding memory accesses.

\paragraph{GPU execution}
The GPU's compute engine contains multiple Processor Clusters (PCs), each containing multiple Streaming Multiprocessors (SMs). SMs are partitioned into multiple processing blocks, each containing specialized processing cores (e.g., INT32 cores), a scheduler and a dispatch unit.
\emph{GPU kernels} to be executed on the GPU are scheduled to SMs and specify the
number of threads to be created. These threads are organised in thread blocks
and grids. Thread blocks are divided into warps. Each warp is a group of
\num{32} parallel threads and gets scheduled by a warp scheduler.

Execution is context-based, where a context represents a collection of resources
and state that is required to execute a GPU kernel. Resources on the GPU are
allocated per context and freed upon its destruction. Each context has its own
assigned address space and allocates at least one channel to receive commands.
However, on current GPUs there is no isolation between contexts that prevents them
from accessing each other's resources.

Modern GPUs have multiple processing pipelines~\cite{nvidia-pipelines} for
different data types. The FMA pipeline executes 32-bit floating point
instructions and integer multiply and add (\texttt{IMAD}). The ALU pipeline executes 32-bit integer, logical, binary, and data movement operations. In addition, there
are also pipelines for 64-bit and 16-bit floating point, and Tensor core
operations. The FMA and ALU pipeline have a separate dispatch port with a warp
dispatch latency of two cycles. Concurrent execution is achieved by alternating
instruction dispatch to different pipelines. The other pipelines use the same
dispatch port and concurrent dispatching to these pipelines is not possible.

\begin{figure}[t]
\centering
\includegraphics[width=0.99\linewidth]{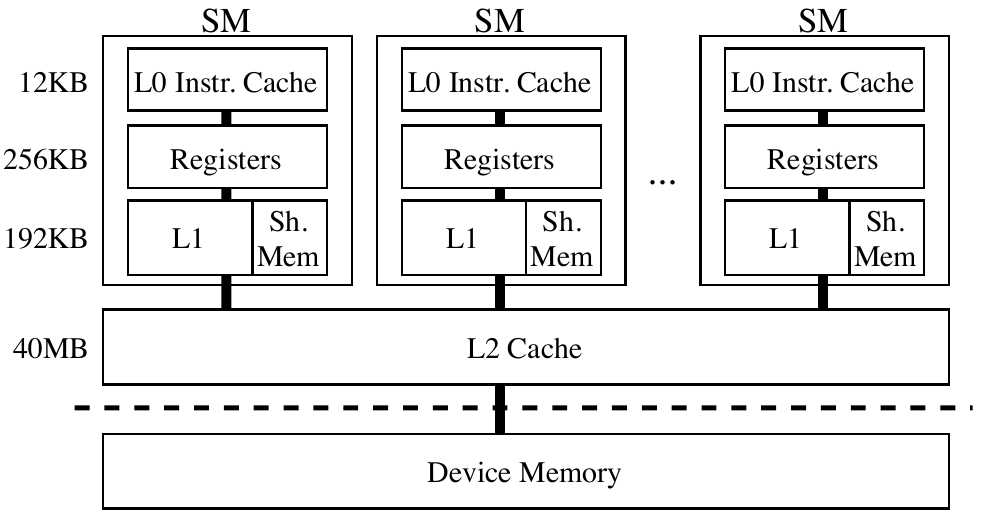}
\caption{Memory hierarchy of a GPU with memory sizes of \nvidia A100 GPU.}
\label{fig:mem_hierarchy}
\end{figure}

\paragraph{GPU memory system}
The memory system on GPUs consists of a memory controller and different memory
levels. The memory levels are associated to the compute system as follows
(see~\cref{fig:mem_hierarchy}). Each processing block includes an L0 instruction
cache and a register file. The combined processing blocks of a SM share a
combined L1 data cache/shared memory that can be partitioned depending on the
workload. Multiple SMs share an L2 cache before pulling data from global
(off-chip) GDDR memory. \emph{Registers} are a shared resource and are allocated among the thread blocks executing on a SM with a granularity of \num{8}. Accessing a register consumes
zero extra clock cycles per instruction, but delays may occur due to register
read-after-write dependencies and register memory bank conflicts. In case a
thread requires more registers than available, the data contained in the registers
is spilled into shared memory. \emph{Shared memory} is not only used for register
spilling, but also enables cooperation between threads in a block. When multiple
threads in a block use the same data from global memory, shared memory can be
used to access the data from global memory only once.

\section{Problem Definition}
In this section, we describe the design goals we strive to achieve, as well as the assumptions and the adversary model we consider.

\subsection{Design Goals}
\label{sec:design-goals}

\paragraph{Verifiable code execution on the GPU}
Verifiable code execution describes the problem in which a verifier wants a
guarantee that some arbitrary code has executed untampered on an untrusted
platform, despite the potential presence of a malicious entity (e.g., malicious
software)~\cite{seshadri2005pioneer}. This problem is typically approached by
verifying code integrity through root of trust attestation, setting up an
untampered code execution environment, and then executing the code.

\paragraph{Data integrity and confidentiality}
In addition to code integrity also the integrity and/or confidentiality of the data executed on the GPU must be ensured. Specifically, we wish to guarantee that the adversary cannot observe or tamper data transferred to/from the GPU by a trusted application that runs in a CPU TEE.

\paragraph{Dynamic root of trust without hardware support}
Dynamic root of trust establishment denotes the problem of dynamically setting
up a trusted computing base (TCB) on an untrusted platform without hardware
support. All code contained in the dynamic root of trust is guaranteed to be
unmodified and it can thus be used to provide externally verifiable code
execution.

\paragraph{Practical to run on a GPU architecture.} The attestation mechanism must be practical to run in a high performance environment with thousands of cores and high memory bandwidth~\cite{keckler2011gpus} while not impacting the performance of the verified program.

\subsection{Assumptions}
\label{sec:assumptions}

\paragraph{Verifier and GPU on the same machine}
We assume that the verifier is executed on the same machine as the GPU we want to
attest. The GPU is directly connected to the host CPU over a bus (e.g.,
PCIe with a latency of \textasciitilde{}\SI{500}{\nano \second}~\cite{neugebauer2018understanding}).

\paragraph{GPU hardware configuration}
We assume that the verifier knows the exact hardware configuration of the GPU,
including the GPU model, the number of cores, the memory architecture, and the GPU clock speed.

\paragraph{Multi-GPU environments}
In heterogeneous multi-GPU environments, we assume that our verification
function runs on the fastest GPU available. In homogeneous multi-GPU systems, an
arbitrary GPU can be selected for computation. Consequently, the adversary
cannot get a performance advantage by running code on other available GPUs.
Furthermore, the dynamic RoT could also be established in sequence (while actively maintaining already established RoTs) starting from the most powerful GPU to the least powerful GPU.

\subsection{Threat Model}
\label{sec:threat-model}
In the following, we discuss the threat model consider by defining the trusted compute base (TCB) and outlining the capabilities of an adversary. The TCB of a system refers to all hardware and software components that are critical to its security, in the sense that bugs or vulnerabilities occurring inside the TCB might jeopardize the security properties of the entire system.

\paragraph{Trusted compute base (TCB)}
We assume an adversary who has complete control over the untrusted host system.
In other words, the adversary has administrative privileges, can tamper with the entire system software, the operating system, or the guest operating system and the hypervisor in case of virtualization.
However, we assume that the hardware primitives of the CPU and GPU, and CPU's firmware are contained in the TCB.
Since \name uses Intel SGX, it inherits the TCB of SGX (which includes the CPU package, trusted libraries, etc.).
In addition, \name{'s} TCB comprises of the user-space GPU runtime and the GPU driver. The runtime is used to program the GPU and transfer data between host and device memory, whereas the GPU driver is responsible for submitting commands to the GPU via the PCI bus and for managing the device. Currently neither the GPU driver nor the GPU runtime are part of SGX. Integrating them into SGX is considered out-of-scope for \name, but could be addressed in future work by porting an open-source driver such as \emph{nouveau}~\cite{nouveau} to SGX. Furthermore, a GPU vendor could also decide to port their proprietary drivers and runtime to SGX.

In addition, we assume that the adversary has physical access to all system hardware, including the GPU. But the adversary cannot perform any physical attacks on the GPU's chip.

\paragraph{Capabilities}
Considering these capabilities, an adversary can read and tamper with code or
data of any victim process, and can access or modify data in DMA buffers or commands submitted to the GPU. Furthermore, the adversary could inject packets in arbitrary locations on the I/O communication path between the host and the GPU. This gives the adversary control over attributes, such as the address of GPU kernels being executed and parameters passed to the kernels. The adversary may also access
device memory directly over MMIO, or map a user's GPU context memory space to a
channel controlled by the adversary. Given the physical access to the system,
the adversary can mount eavesdropping attacks on the host memory and PCIe bus.
In GPUs that support multi-tasking, malicious kernels can be dispatched to the
GPU, thereby accessing memory belonging to a victim's GPU context. Since the
adversary controls the hypervisor and thus the mapping between VMs and virtual
devices, these attacks are possible even in a virtualized environment.

\paragraph{Out of scope}
Since this work tackles the problem of trusted execution on the GPU, we do not
consider attacks that target the defense of SGX, such as physical attacks to the CPU package or side-channel attacks on SGX. In addition, we do not consider system availability attacks that prevent the execution of our process, as an adversary with the described capabilities can always prevent the deployment of computing tasks on the GPU.

\section{\name Overview}
\label{sec:design}

\begin{figure}[t]
\centering
\includegraphics[width=0.99\linewidth]{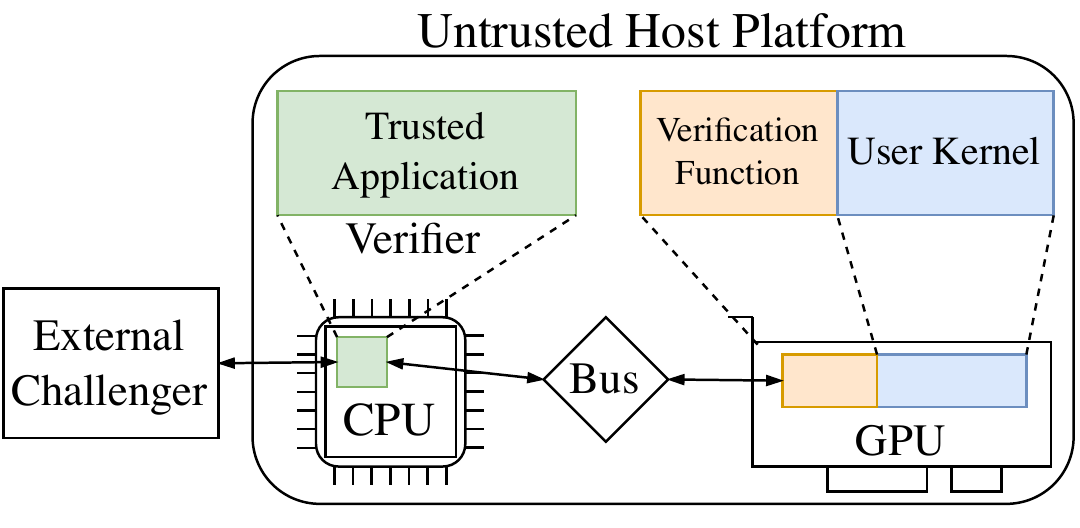}
\caption{Abstract system model.}
\label{fig:system_model}
\end{figure}

\name addresses the problem of verifiable code execution on a GPU without hardware
support, in which the verifier wants a guarantee that some arbitrary code (i.e., the user kernel) has executed untampered on an untrusted GPU platform, even in the presence of an
adversary. \Cref{fig:system_model} illustrates the abstract system model we consider.

\name comprises of two main components. The first component is the verifier, which runs as a trusted
application on the host CPU (e.g., using Intel SGX~\cite{costan2016intel}) and is attested by an external challenger.
The second component is the verification function (VF), which runs on the
the untrusted GPU. The VF computes a checksum over its own code, and is constructed in an intricate way such that if a change is applied to the VF then either the execution will slow down in an externally detectable manner, or the checksum value will be incorrect.

The verifier dispatches to the GPU the VF and then invokes it repeatedly with a series of challenges while measuring the VF execution time for each invocation. For
every challenge, the VF computes a checksum value and returns it to the verifier. Using the same VF logic, the verifier independently computes and verifies the correctness of the checksum value. If the checksum returned by the VF is correct \emph{and} it is returned within the expected time, the verifier obtains a guarantee that a dynamic root of trust on the GPU was established.

\begin{figure}[t]
\centering
\includegraphics[width=0.7\linewidth]{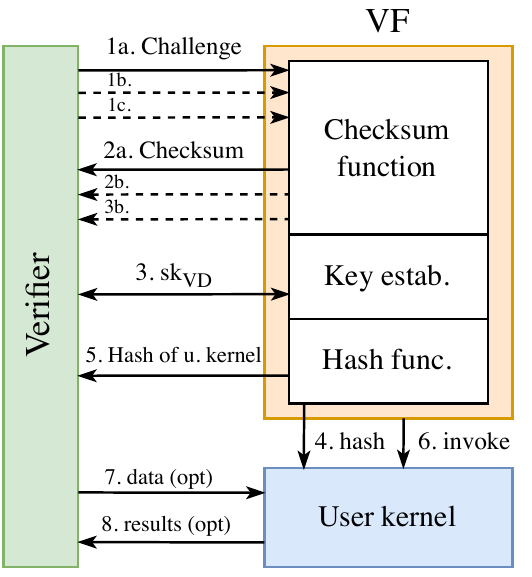}
\caption{Overview of \name. The numbers represent temporal ordering of events.}
\label{fig:overview}
\end{figure}

Once the dynamic root of trust has been established, the VF checks the
integrity of the user kernel, sets up an untampered execution environment, and finally runs the user kernel. During the setup of the execution environment, a shared key
between the verifier enclave and GPU is established; afterwards, only commands
authenticated with this key are accepted, including the movement of (encrypted if needed) data between host and GPU. \Cref{fig:overview} shows an overview of \name including a sequence of events.

\section{Verification Function (VF)}

The VF that runs on the untrusted GPU is the fundamental component
of \name.
% It is responsible for computing a checksum value, setting up an untampered execution environment and measuring the integrity of the user kernel prior to its execution.
We now describe in detail these tasks and the challenges that they entail.

\subsection{Design Requirements}

The VF must be carefully constructed in such a way that if an adversary were to tamper with the VF or the user kernel, it would result in either a wrong checksum or a noticeable time delay.
Before offering a concrete design for the VF, we describe several required properties and outline how these properties influence the correctness of the checksum or the VF execution time.
We defer our security analysis to~\cref{sec:data-sub-attack}; the following properties also account for the attack surface analyzed therein.

\begin{enumerate}[nolistsep]
    \item \emph{Time-optimal implementation.} The implementation of the VF must be \emph{time-optimal}. Otherwise, the adversary could use a faster implementation and use the time saved to forge the checksum (e.g., by injecting instructions).

    \item \emph{Maximize resource usage during checksum computation.}
    To prevent the adversary from running any other computation during the checksum computation, the VF should maximize its resource usage on the GPU by using all available SMs and avoiding ``empty'' threads. Moreover, each thread should use the maximum number of available registers to prevent the adversary from using those registers. Thus, if the (tampered) computation attempts to use more registers than available, the values of the affected registers are spilled into shared memory, resulting in a noticeable execution time difference (4- vs. 30-cycle latency for registers and shared memory, resp.).

    \item \emph{Predictable execution time.} The execution on GPUs is optimized to achieve high data throughput with determinate latency, but the execution time is non-deterministic (e.g., due to multi-threaded execution, scheduling, and caching). The VF execution time should have low variance so that the verifier can predictably determine the correct execution time.

    \item \emph{Challenge-dependent checksums.} To prevent the adversary from pre-computing the checksum before making changes to the VF, and to prevent the replay of old checksum values, the checksum needs to depend on an unpredictable challenge sent by the verifier.

    \item \emph{Strongly-ordered checksum code.} A strongly-ordered function is a function whose output differs with high probability if the operations are evaluated in a different order.
    If the adversary attempts to gain an advantage by altering the order in which single instructions or entire sequences of instructions are executed, the verifier should receive a wrong checksum or record a noticeable increased execution time.

    \item \emph{Include architectural complexity.} Subtle data attacks need to be prevented (more details in~\cref{sec:data-sub-attack}). For instance, the adversary can keep a correct copy of any memory location in the tampered VF and when the checksum computation attempts to read one of the modified memory locations, the read is redirected to the location where the adversary has stored the correct copy. To maximize the time overhead for such attacks, the checksum computation should include some architectural complexity (e.g., a pseudo-random memory access).
\end{enumerate}

\subsection{Concrete VF Design}
The VF consists of four parts: 1) initialization, 2) self-verifying
checksum function to establish a dynamic root-of-trust, 3) establishing
an untampered execution environment (including a key establishment
protocol between the verifier and the GPU), and 4) a hash function to measure the integrity of the user kernel. We describe each in the following.

\subsubsection{Initialization of the VF}
\label{sec:initVF}
During the initialization phase, the verifier dispatches the VF code to the GPU.
For this purpose the GPU first allocates a memory buffer and returns the
buffer's base address to the verifier. The code of the VF is
then copied into the buffer at specific offsets.

\subsubsection{Self-Verifying Checksum Function}
The checksum function is used to obtain a guarantee that the integrity of the VF code running on
the GPU is unaffected by an adversary. For this purpose, the
checksum function computes a checksum over the entire VF code. The resulting
checksum can be used as a \emph{fingerprint} of the VF and enables detection of changes
to the VF code. If an adversary modifies the VF code, the checksum will
differ with high probability. Thus, once the verifier receives a correct
checksum within a threshold time, it has a guarantee that the VF code running on the GPU is unmodified.

Since the checksum computation code is part of the VF and will thus be
included in the checksum calculation, the checksum function
computes the checksum over its own instruction sequence and
verifies itself. This property is further referred to as \emph{self-verification}.

\paragraph{Checksum initialization}
GPUs contain multiple multiprocessors that can be used for parallel execution.
To exhibit the maximal computational power of a GPU, the verifier sends a set of
challenges containing a specific challenge value for each multiprocessor. Upon receiving a set of challenges, each multiprocessor uses its challenge as a seed value to initialize all per-thread state with random data and a pseudo-random number generator (PRNG) that is used during the checksum computation.
Each thread has its own set of registers which are used to store the running checksum values and a data
pointer. The data pointer references the VF code in the initially allocated
buffer (\cref{sec:initVF}).

\paragraph{Self-modifying code}
In memory copy attacks, the adversary replaces the checksum function with an altered version and executes it (described in detail in~\cref{sec:mem-cpy-attack}).
To prevent memory copy attacks, the execution state -- namely, the program counter (PC)
and data pointer (DP) -- needs to be included in the computation of the checksum.

The GPU maintains execution state per thread, including a program counter and call stack, however, does not expose it to the high-level programming models. In the low-level programming models, the current PC could be loaded into a register using instruction patching (\texttt{LEPC} instruction, see~\cref{sec:instr-generation}).
However, loading the PC into a register and including it into the checksum computation does not prevent memory copy attacks.
An adversary could replace the instruction used to load the PC with a move instruction with an constant immediate value without causing any computational overhead. An alternative to directly loading the PC to prevent memory copy attacks could employ boundary checks (e.g.,
the DP needs to be in close proximity to the PC), but could be circumvented by an attacker in the same way.

Instead of directly including the PC in the checksum computation, we use \emph{self-modifying code}, such that the execution of these instructions depends on the current value of the checksum. If the adversary wants to successfully execute a memory copy attack, he would need to closely monitor the execution of the checksum function and modify the instruction in its copy accordingly. This manipulation of the code in two location causes a constant time overhead in each execution of the self-modifying code.

\paragraph{Checksum loop}
The checksum computation is performed iteratively where each iteration contains the following steps. Each iteration executes the same number and type of instructions and has a constant
execution time. However, the self-modifying code depends on the current checksum value and thus changes in each iteration.
\begin{enumerate}[nolistsep]
  \item \emph{Pseudo-random memory access of VF code}. In data substitution
attacks (see~\cref{sec:data-sub-attack} for details) the adversary keeps a
correct copy of any memory location in the VF code it modifies and attempts to
redirect all memory reads that access the modified memory location with reads of
the correct copy it has stored. Thus, the adversary's checksum result will be
correct despite the modification of the VF. To maximize the adversary's time
overhead for this attack, the checksum code reads the memory region
containing the VF code in a pseudo-random pattern. This pseudo-random memory
access prevents the adversary from predicting which memory-resident instruction will read the
potentially-modified memory location and forces the adversary to monitor every
memory read by the checksum code.

  \item \emph{Update the checksum}. The running checksum values are updated to
include the accessed VF code into the checksum value using a sequence of
instructions. To achieve a time-optimal
implementation, we use simple arithmetic and logical instructions (e.g.,
\texttt{add}, \texttt{<{}<}, \texttt{>{}>}, etc.) that are challenging to
implement faster or with fewer operations. Taking inspiration from the strong ordering used in~\cite{seshadri2005pioneer}, the instructions used to update the checksum
alternate between arithmetic and logical instructions to enforce a strong
ordering of the instructions.
%After updating the checksum with these instructions, we rotate the checksum value using shift instructions (with carry over) by a varying prime number of positions (e.g., 3, 7, 13, or 19).
%This changes the position of checksum bits in each iteration of the checksum loop.
Including the pseudo-randomly accessed VF code segment enables self-verification of the
checksum function.% and also prevents the adversary from hardcoding the next jump target.

  \item \emph{Include the data pointer}. In the next step, the data pointer that is
kept as a running value in each thread is included into the checksum. This
ensures that the memory region is untampered and prevents memory copy attacks
that tamper the VF, but keep a copy of the original VF in a different memory location
(see~\cref{sec:mem-cpy-attack}).

%   \item \emph{Compute the next jump target}. The next jump target is determined
% using the least-significant bits of the current checksum value. Because these
% code blocks are fixed size and allocated consecutively, these bits can be used
% as an index for the next code block.

%   \item \emph{Include the jump target address into the checksum computation}.
% Then, the jump target address is incorporated into the checksum computation to
% ensure a strong ordering between the executed checksum code blocks.

%   \item \emph{Jump to the next block}. The last step of a checksum block is
%   a jump based on the target address generated earlier.

  \item \emph{Self-modifying code}. The instructions of the self-modifying code depend on current value of the checksum function and are changed in each iteration of the checksum function. In our case the current value of the checksum function is used as an immediate value for an instruction (see \cref{sec:sage-impl} for details).

\end{enumerate}
If an adversary alters the checksum function but wants to forge a correct checksum output,
he has to manipulate the values of one or more of the inputs in every iteration of the
checksum code, causing a constant time overhead per iteration.

\paragraph{Checksum epilog}
Since the checksum computation is conducted using individual threads located on different
multiprocessors, the checksum values need to be aggregated before sending the checksum
result back to the verifier. This aggregation is conducted in three steps.
%(as illustrated in~\cref{fig:checksum_finalization}).
First, we aggregate the checksum per warp. Each of the per-thread checksums is added pairwise to obtain a warp-level checksum. Second, the warp-level checksums are aggregated by thread block using shared memory. Finally, we aggregate the checksum per grid using global memory. Each of the aggregation steps uses a pairwise addition (which is mapped to an atomic add instruction in native assembly). The final result of the checksum computation is then sent to the verifier.

\begin{figure}[h]
% \vspace{-0.1cm}
\centering
\includegraphics[width=0.95\linewidth]{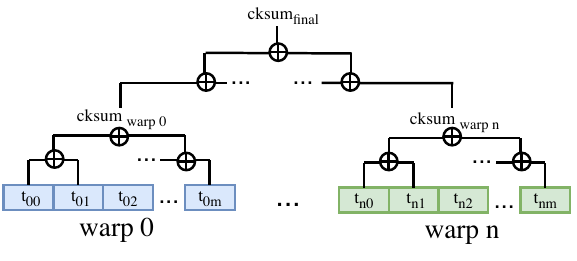}
% \vspace{-0.3cm}
\caption{Aggregation of the intermediate checksum values into the checksum result.}
\label{fig:checksum_finalization}
% \vspace{-0.2cm}
\end{figure}

\subsubsection{Untampered Execution Environment}
After establishing a dynamic root-of-trust on the device, the VF sets up an execution environment in which the user kernel is guaranteed to
run untampered. This includes setting up a shared secret between the verifier
and the device, and checking the authenticity of the user kernel to be executed on the
GPU using a hash function. The shared secret can then be used to authenticate
and encrypt commands and data sent by the verifier to the device and vice versa.

\paragraph{Key establishment}
To establish a shared secret between the verifier and the device, we rely on the
SAKE protocol~\cite{seshadri2008sake}, a protocol for key establishment between
neighboring nodes in sensor networks without requiring any prior secrets. The
protocol is based on the Diffie-Hellman key exchange protocol and uses the Guy
Fawkes protocol~\cite{anderson1998new} for authentication. The Guy Fawkes
protocol is based on hash chains and relies on the property that each of the
participants needs to authenticate the other party's hash chain. In SAKE, this
authentication is achieved using software-based attestation and exploits the
asymmetry in the checksum value and the computing time between the genuine
checksum function and a modified checksum function. This allows us to use the
resulting checksum as a short-lived ``secret.'' Furthermore, the SAKE protocol
assumes that the adversary does not introduce any computationally powerful nodes into
the network, which aligns with the assumptions for \name (see~\cref{sec:assumptions}).
% 1) we assume that we run on the fastest available GPU. 2) remote compute cluster attacks would
% introduce additional latency

% SAKE assumptions:
% - node configuration is identical
% - silicon ID -> unique, immutable identifier
% - hardware source for randomness
% - all messages are reliably delivered

To apply the SAKE protocol to \name, we change the protocol as follows: 1) The
checksum function in SAKE that was proposed for the use in sensor networks is
replaced with \name{'s} checksum function. 2) Instead of both participants
acting as challengers, only the host enclave will engage as a challenger.
% 3) Since both the host enclave and the GPU are located in local environment the silicon ID check is not required and thus omitted.
3) We replace the cryptographic primitives used in the
protocol with AES-CMAC as the MAC function and SHA256 as the hash function.

Assuming the changes above, the key establishment protocol in \name works as
follows. First, the verifier sets up its own hash chain and DH public key as:
\begin{equation}
\textstyle V: \enspace v_0 = g^a \text{ mod } p \enspace v_1 = H(v_0) \enspace v_2 = H(v_1)
\end{equation}
where $a$ is a random bitstring $a \gets^\R \{0,1\}^n$
Then, it sends $v_2$ to the device and records the current time as $t_0$.
\begin{equation}
\textstyle [t_0] \quad \mess{V}{D}{v_2}
\end{equation}
Upon receiving $v_2$, the device uses it as a challenge for the checksum function
and then uses the computed checksum and a random value to generate its own hash
chain and replies to the verifier:
\begin{equation}
\textstyle D: \enspace w_0 = H(c \fdelim r) \enspace w_1 = H(w_0) \enspace w_2 = H(w_1)
\end{equation}
where $r$ is a random bitstring $r \gets^\R \{0,1\}^n$, $c$ is the result of the checksum computation and \fdelim refers to concatenation.
\begin{equation}
\textstyle [t_1] \quad \mess{D}{V}{w_2, \ \text{MAC}_c(w_2)}
\end{equation}
The verifier checks if the measured execution time ($t_1 - t_0$) matches the
expected execution time and aborts the protocol otherwise. In the meantime, the
device sets up its own DH public key:
\begin{equation}
\textstyle D: \quad b \gets^\R \{0,1\}^n \quad k = g^b \text{ mod } p
\end{equation}
Then, the verifier and the device gradually disclose the remaining of their hash
chains to each other:
\begin{align}
\mess{V}{D}{v_1}&\qquad
\mess{D}{V}{w_1, \ k, \ \text{MAC}_{w_2}(k)}\\
\mess{V}{D}{v_0}&\qquad
\mess{D}{V}{w_0}
\end{align}
For each message the recipient checks whether the received value matches the
expected hash chain. Finally, the verifier $V$ and the device $D$ compute the shared
secret key $sk_{VD}$:
\begin{equation}
\textstyle sk_{VD}  = k^a = (g^b)^a \text{ mod } p \quad sk_{VD}  = v_0^b = (g^a)^b \text{ mod } p
\end{equation}

\paragraph{Authenticity check of user kernel}
To check the authenticity of the user kernel, the VF uses a
hash function $H$ (SHA256) and computes a hash over the user kernel located in
memory on the device concatenated with a random value $r$ provided by the verifier.
\begin{equation}
\textstyle h = H(r \fdelim \text{code})
\end{equation}
Then, it returns the hash value $h$ to the verifier that checks whether it matches the expected hash value.

\subsubsection{Protected Data Transfer to the GPU}
After the dynamic RoT has been established on the GPU and the integrity of the user kernel has been checked, the host enclave can start transferring data that will be processed by the user kernel to the GPU.
Depending on the sensitivity and security criticality of the domains, the data could be either \emph{authenticated}
and/or \emph{encrypted} using the established symmetric key $sk_{VD}$. For authenticity, the data transfer can already be started while the host enclave is checking the result of the checksum computation. However, if the
data must remain confidential, the data transfer must be deferred until the checksum returned by the untrusted device is considered valid.

\section{Implementation}
To establish the dynamic RoT on the untrusted GPU, \name relies on the checksum result, but also measures the execution time of the VF to assess its integrity. This requires a \emph{time-optimal implementation} of the VF. Otherwise, the adversary could use a faster implementation and use the time saved to forge the checksum (e.g., by injecting instructions). The requirements to achieve a time-optimal implementation on the Ampere architecture (further discussed in~\cref{sec:time-opt-requirements}) include maximizing GPU utilization, consume all available compute resources, optimally filling the processing pipelines, and optimize cache usage.

Unlike the higher levels of the CUDA computing platform such as the CUDA C/C++ language extension and the parallel thread execution (PTX) virtual machine and instruction set architecture, \nvidia provides very little information about the hardware specific instruction sets for a specific target architecture. Moreover, even if one resorts to write inline PTX virtual assembly, the Streaming (or Shader) Assembler (SASS) code emitted by the compiler often does not achieve the performance of native GPU applications. The execution of microcode that has been compiled using the regular CUDA compiler often is in the order of 10x slower compared to optimized microcode~\cite{jia2019dissecting, jia2018dissecting}.
As a consequence, libraries used for high-performance computing (e.g., cuBLAS~\cite{cublas}) contain highly optimized microcode tailored to a specific architecture.
In addition to the performance gap to native GPU code, the user has no control over the translation from PTX virtual assembly to the SASS assembly for the target architecture.

\begin{figure}[t]
\centering
\includegraphics[width=0.9\linewidth]{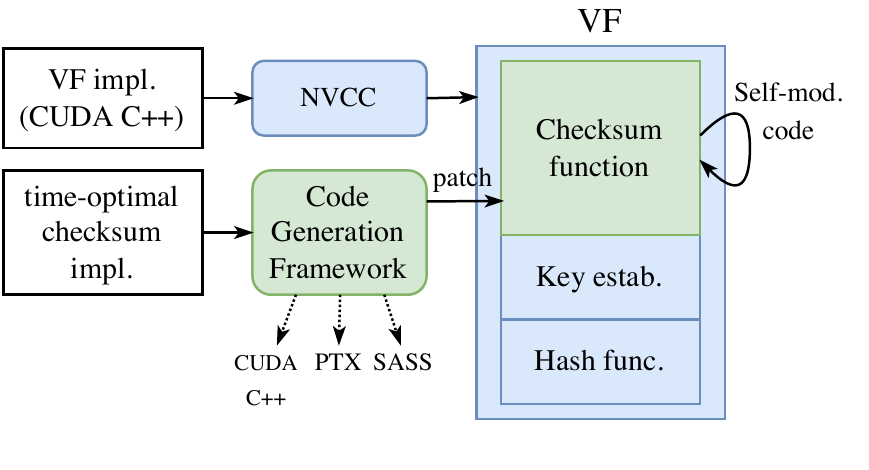}
\caption{Code pipeline to generate the VF microcode. The green blocks refer to the optimized microcode generated using our code generation framework.}
\vspace*{-3mm}
\label{fig:code_pipeline}
\end{figure}

To achieve a time-optimal implementation, we are required to implement a custom instruction generation framework that allows patching of binary microcode with a highly optimized version. The implementation of this framework requires understanding the Ampere architecture and the instruction format used in microcode. \Cref{fig:code_pipeline} illustrates the pipeline used to generate the VF. The VF is implemented using CUDA C++ and compiled using \texttt{NVCC}. However, the section containing the checksum function is patched using an optimized implementation generated as binary microcode using our framework.
Our framework and implementation of \name can be found here \url{https://github.com/spcl/sage}.

\subsection{Instruction Decoding}
To understand the instruction format used in the recent Ampere GPU architectures, we implemented a framework that allows decoding of instructions using \texttt{cuobjdump} and \texttt{nvdisasm}~\cite{nvidia-utils} by decoding handcrafted code samples and samples from existing CUDA libraries (e.g., cuBLAS~\cite{cublas}).

\begin{figure}[h]
\centering
\vspace*{-5mm}
\includegraphics[width=\linewidth]{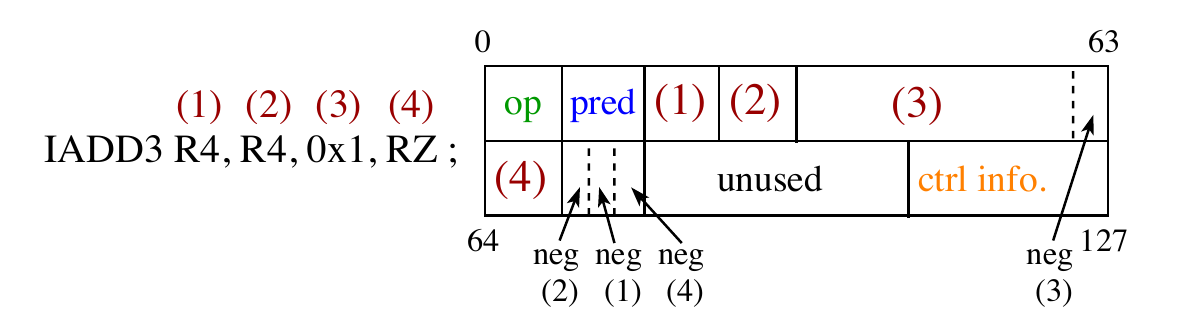}
\caption{Instruction as decoded by \texttt{nvdisasm} and its format. \texttt{pred} denotes predicates, \texttt{op} refers to the operation code, and \texttt{neg} allows negating the corresponding parameter.}
\vspace*{-1mm}
\label{fig:instruction-decoding}
\end{figure}

\paragraph{Instruction format}
%The instruction encoding format used in \nvidia GPUs is adopted across the most recent GPU architectures.
NVIDIA's Ampere architecture adopts the same general instruction format as its predecessors Turing and Volta~\cite{jia2018dissecting, jia2019dissecting}.
All these architectures use 128 bits to encode both an instruction and its associated scheduling control information. The encoding that is used in these architectures is fixed length and uses similar encodings for all instructions.
Figure~\ref{fig:instruction-decoding} illustrates a typical instruction encoding.

\paragraph{Control information}
The control information section in the instruction encodes scheduling decisions taken by the compiler that the hardware must enforce. The control information is organized as follows: reuse flags (\SI{4}{\b}), wait barrier mask (\SI{6}{\b}), read barrier index (\SI{3}{\b}), write barrier index (\SI{3}{\b}), yield flag (\SI{1}{\b}), and the number of stall cycles (\SI{4}{\b}).
The reuse flags allow data reuse between instructions without accessing any register ports. The wait barrier mask and indices are used for instructions with variable latency (e.g., instructions involving a memory access). These dependency barriers can be used to enforce the completion of variable-latency instructions. The yield flag is used to balance the workload assigned to a processing block. The stall cycles indicate the latency of the instruction before issuing the next instruction. Jia et al.\ present a detailed description of the control information~\cite{jia2019dissecting}.

\subsection{Instruction Generation}
\label{sec:instr-generation}
Understanding the instruction format allows us to generate the specific instructions we need for our implementation. These instructions then need to be translated to the correct binary format. For this purpose, we implement an instruction generation framework that allows emitting instructions either in CUDA C++, the virtual assembly language PTX, or as binary microcode that is natively executed on the GPU.

The instruction can be defined in the following format, where the section separated using pipes describes the control information for the instruction (barrier mask B, read barrier index R, write barrier index W, yield flag Y, and number of stall cycles S):\\
\texttt{\small B......|R.|W.|Y1|S1| IMAD.U32 R28, R28, 2048, R28;}

Our instruction generation framework then translates the instruction to the selected target language (CUDA C++, PTX, microcode). This allows us to rapidly prototype checksum functions and compare performance between implementations in each of the languages.

\subsection{Time-optimal Technical Requirements}
\label{sec:time-opt-requirements}
We formulate the following technical requirements for a time-optimal implementation of the checksum function. These are subject to characteristics of the target architecture; in our case, the \nvidia Ampere architecture.

\paragraph{Maximize resource consumption}
To maximize the resource consumption during the checksum computation, the checksum function must use all available compute resources.
The \nvidia A100 GPU has 108 Streaming Multiprocessors (SMs) each containing \num{64} \texttt{FP32} and \num{64} \texttt{INT32} units~\cite{nvidia-ampere} that must be used during each clock cycle.

\paragraph{Optimally fill FMA and ALU pipelines}
Since both the FMA and ALU pipelines have an instruction issuing latency of \num{2} clock cycles, \texttt{FP32} and \texttt{INT32} instructions must be interleaved to fully saturate both pipelines. In addition, instructions that use registers with a direct dependency must be executed with a latency of at least \num{4} clock cycles to avoid pipeline stalls (e.g., read-after-write dependency).

\paragraph{Optimal GPU utilization}
To achieve full GPU utilization, the number of threads per thread block needs to be picked according to the target architecture.
% For Ampere, \num{32} blocks per thread result in an optimal utilization~\cite{nvidia-occupancy}.
The A100 achieves full GPU occupancy by assigning 2 blocks of size 1024 to all the 108 available SMs (216 total).
Each SM has 65,536 32-bit registers available for threads. To use all registers during the checksum computation while maintaining full utilization of the GPU, \num{32} registers are assigned per thread~\cite{nvidia-occupancy}.

\paragraph{Cache size} The code blocks should not exceed the capacity of L0 and L1 instruction caches.

\subsection{Selection of Optimal Overheads}
\label{sec:selection-of-optimal-overheads}
An optimal implementation of a checksum function should perform a useful computation step in each clock cycle. In practice, this requires a highly optimized use of the underlying hardware. In the following, we show a recipe for building such a checksum function for the A100 GPU.

Unutilized clock cycles are mainly caused by instruction cache misses, global memory access latency, pipeline stalls, and jumps to the beginning of loops. In the beginning of each clock cycle, the SM warp scheduler selects a subset of warps (up to 4 on A100) from all active warps (up to 64 on A100) to execute. This selection mechanism can avoid performance losses if at least 4 are ready to execute on each clock cycle. % The remaining active warps can wait due to various delays.

To analyze the performance of the checksum function, we use a simplified model of the number of clock cycles per instruction. We distinguish the total number of useful clock cycles \texttt{X} and overhead cycles \texttt{Y}, so that the total number of clock cycles spent by the code using a single thread is \texttt{X+Y}. For example, with proper instruction ordering to avoid pipeline stalls, an \texttt{IMAD} instruction has \texttt{X=1} and \texttt{Y=0}. An instruction reading from global memory has \texttt{X=1} and approximately \texttt{Y=250}. To prevent attacks on the checksum function, the value \texttt{Y} must not exceed \texttt{X(64/4 - 1)}. Then, the GPU scheduler will be able to completely hide the overhead \texttt{Y} so that the actual amount of time spent will be \texttt{X}.

Integer shifts and multiplications with addition directly affect the result of the checksum calculation. However, among the useful instructions in \texttt{X}, some do not affect the checksum. One of them is the instruction for jumping from the end of the loop body to its beginning. The attacker may try to unroll a few iterations of the loop to save the clock cycles required to perform this jump (and potentially misuse them for an attack). To prevent such attacks, we unroll the loops until it is not possible to unroll them further without causing instruction cache misses. The target value \texttt{Y} for unrolling must be so large that one additional instruction cache miss will increase it to \texttt{Y'} such that the hardware scheduler can no longer hide. A similar attack and defense against it applies to other instructions inside the body of the checksum function.

In practice, we have noticed that achieving this level of control over the order of instructions, and the arrangement of unrolled loops is very difficult without vendor support: the documentation on SASS and hardware details is deliberately kept closed ease backward-compatibility hassles. It is especially difficult to control instruction cache misses because of the use of self-modifying code to protect against memory copy attacks. The only way to invalidate the instruction cache on the A100 is to overflow it with the block of instructions of the cache size, so controlling the value of \texttt{Y} by changing the size of the checksum function is not possible. That leaves only memory accesses and jumps that can change \texttt{Y}. We assume that adding an instruction to invalidate the instruction cache requires minimal (or no) changes to the GPU architecture because a similar instruction already exists for the data cache (\texttt{discard} in PTX ISA or \texttt{CCTL} in SASS).

\subsection{Implementation of \name}
\label{sec:sage-impl}
\paragraph{Verifier}
We implement the verifier enclave using the Intel SGX SDK~\cite{linux-sgx} and its \texttt{tcrypto} library~\cite{linux-sgx-tcrypto}. The enclave creates a CUDA context on the GPU, loads the VF as a module, and calls the VF kernel. To generate nonces in the enclave that are then transferred to the GPU as challenges, we use AES-CTR with an IV that has been generated using a TRNG during the enclave creation.

\paragraph{VF}
The VF is implemented in CUDA C++, except the checksum function component, which is patched by binary microcode using our framework.
The checksum function executes a loop containing the following steps:
\begin{enumerate}[nolistsep]
    \item The iteration counter is increased and checked if the maximum number of iterations is reached.
    % \item The next jump target $T$ is computed based on current checksum value $C$ as
    % $T = \text{block\_offset} + C \ \% \ \text{\# blocks} \times \text{block\_size}$.
    %     %\begin{equation}
    %     %    T = \text{block\_offset} + C \ \% \ \text{\# blocks} \cdot \text{block\_size}
    %     %\end{equation}
    %     To optimize this computation (and avoid using a division), we pad the function blocks with \texttt{NOP} instructions to a block size that is a power of \num{2}. This allows to use an \texttt{AND} instruction to find the next jump target. Furthermore, all threads involved in the computation have the same jump target to avoid thread divergence.
    \item A data block $D$ of the VF is read from memory and will be included into the checksum computation as
    $D = \text{data\_ptr} + 4 \times C \ \% \ \text{data\_size}$.
        %\begin{equation}
        %    D = \text{data\_ptr} + 4 \cdot C \ \% \ \text{data\_size}
        %\end{equation}
    The read from main memory takes \textasciitilde{}\numrange{250}{500} cycles to be completed. The GPU compiler sets a read barrier for this instruction and the GPU stalls the compute pipeline until the read has been completed. However, stalling the checksum computation would allow the adversary to execute its own instructions.
    \item Instead of stalling, we design a pattern of instructions that is executed while waiting for the memory read to be completed (``busy waiting''). This pattern must fully utilize both the FMA and ALU pipeline to which instructions are dispatched alternatingly. To achieve this utilization we use a \emph{shift-and-add} pattern that can be implemented using a single instruction on both the FMA and ALU pipeline (\texttt{IMAD} and \texttt{LEA.HI}). As a consequence, the previous access to global memory becomes ``invisible'' in terms of latency.
    \item Once the memory load is completed, the checksum is updated using a thread block-specific computation. The block specific computation consists of alternating left or right shifts with addition. Each instruction uses arbitrarily chosen shift size to make sequences of such shifts unique. Then, the current iteration index and the value of unused registers during the computation are incorporated into the checksum.
    \item After updating the checksum function, we compute the self-modifying code that consists of the following binary instruction:
    \texttt{x+=x>{}>N},
    where the immediate \texttt{N} depends on the current checksum value. We overwrite immediate parameter with the current value of the checksum. Thus, the value of $N$ changes for each iteration and ensures that we are executing the code that we are verifying. To avoid race conditions when updating the immediate value of these instructions, these instructions are required to be located in different memory areas for each thread block.
    % \item Finally, we jump to the next code block based on the previously computed jump target using a branch instruction.
\end{enumerate}

\paragraph{Key establishment}
For the key establishment protocol based on the modified SAKE protocol, the GPU needs to be able to generate random values.
Given that the adversary knows the entire code executing on the GPU, we cannot use a secret provided by verifier to initialize the pseudo-random number generator used in the protocol, but instead must rely on a true random number generator.

\subsection{Random Number Generation on GPUs}
There are two main categories of random number generators: pseudo-random number
generators (PRNG) and true random number generators (TRNG). PRNGs are
deterministic, producing the same set of random numbers for a particular input
seed, while TRNGs are non-deterministic and typically use a physical source of
randomness, whereby every run will produce truly unpredictable random numbers.
In \name, we require a TRNG implementation to generate keys and seed values for
the PRNG that will be used in the key establishment protocol. For the PRNG, we
utilize the CUDA random number generation library (cuRAND)~\cite{curand}.

\paragraph{TRNG implementation on GPUs}
\label{sec:gpu-trng}
Approaches that use physical unclonable functions (PUFs) to initialize PRNGs on the GPU~\cite{g-puf, van2015investigating, decay-puf} are not practical to be used in \name as they either require resetting the GPU or use features that are under control of the
adversary (e.g., voltage supplied to the GPU).
Consequently, we use a TRNG implementation is based on race conditions in multi-core environments caused by simultaneous memory accesses to shared variables and takes advantage of uncertainties that arise when cores simultaneously access a particular memory location~\cite{teh2015gpus}. In our case, each simultaneous memory access unpredictably flips bits stored in shared variables. This unpredictability enables the GPU to generate noise which can be sampled and then used as an entropy source. We evaluated our implementation using statistical tests such as
NIST SP 800-22~\cite{smid2010statistical}, DIEHARD~\cite{marsaglia1996diehard},
and ENT~\cite{ent}. The TRNG implementation passes all standard tests and
achieves a throughput of \SI{4}{\kBps} on \nvidia A100 GPUs and thus takes around \SI{8}{\milli \second} to generate an output of \SI{256}{\bit s}. The TRNG provides \SI[round-mode=places,round-precision=6]{7.999996}{\bit s} of entropy per byte (measured using ENT~\cite{ent}).

\section{Evaluation}
\label{sec:evaluation}
We evaluate the practicality of our approach by evaluating its performance on the \nvidia A100 GPU and show overheads introduced by the VF compared to a regular execution of the user kernel. Finally, we evaluate the robustness of the VF to potential attacks and show that any additional instructions will result in a detectable overhead.

\paragraph{Evaluation setup}
To evaluate the performance of the checksum function, we use an evaluation setup based on an ASUS RS720-E10-RS12E equipped with a Intel Xeon Gold 6348 CPU~\cite{eval-cpu} which natively supports SGX instructions (further referred to as \emph{Intel}), and a \nvidia A100 GPU. We run the SGX enclave in both native and simulation mode. To benchmark the execution time of the verification process, we also run the VF on a dual-socket system with an AMD EPYC 7742 CPU (further referred to as \emph{AMD}).

\paragraph{Register consumption}
For the execution of the checksum function, the loop counter, data pointer, and the checksum result are stored in registers. In addition to those registers, we use 22 additional registers to store intermediate state during the computation of the checksum. In total, the checksum function verifies 524,288 bytes. The beginning of the buffer contains the checksum function itself, whereas the remainder is filled with pseudo-randomly generated values.

\begin{table}[h!]
\small
\centering
\begin{tabular}{@{}lrrrr@{}}
    \toprule
    Experiment Nr. & 1 & 2 & 3 & 4 \\
    \midrule
    self-modifying code & \xmark & \xmark & \cmark & \cmark \\
    instructions & 428 & 429 & 8,342 & 8,342 \\
    iterations & 100,000 & 100,000 & 1,000 & 1,000 \\
    inner iterations & 0 & 0 & 0 & 5000 \\
    inner instructions & 0 & 0 & 0 & 216 \\
    \midrule
    verification (AMD) [s] & 21.6 & 21.6 & 9.99 & 497 \\
    verification (Intel) [s] & 102 & 102 & 47.0 & 2337 \\
    \midrule
    runtime $T_{avg}$ [s] & 0.4941 & 0.4977 & 0.1309 & 12.40 \\
    \% of GPU peak perf. & 99 & 98 & 75 & 100 \\
    \midrule
    adversarial NOP & \xmark & \cmark & \xmark & \xmark \\
    runtime $\sigma$ [s] & 0.0009 & --- & --- & --- \\
    runtime $T_{min}$ [s] & --- & 0.4966 & --- & --- \\
    $T_{avg} + 2.5\sigma$ [s] & 0.4964 & --- & --- & --- \\
    \bottomrule
\end{tabular}
\caption{Evaluation of checksum implementations.}
\label{table:checksum_eval}
\end{table}

\paragraph{Summary of results}
\Cref{table:checksum_eval} summarizes our experiment series conducted to evaluate the performance of \name{'s} VF.
We distinguish between two categories depending on whether the checksum function contains self-modifying code or not.
Depending on the category, the total number of instructions and number of checksum loop iterations are adapted.
For each experiment, we report the VF's execution time on the GPU, the utilization ratio during the checksum execution,
the verification time on the CPU, detection threshold, etc. 

The first experiment demonstrates our best reference implementation. The second experiment simulates an attack on the checksum function from the first experiment. In the third experiment, we show the effect on the performance of adding self-modifying code to the reference implementation. The fourth experiment shows a possible technique to compensate for the loss of performance with enabled self-modifying code.

\subsection{VF Performance}
\label{sec:vf-perf}
To evaluate the performance of \name{'s} VF, we report its average runtime and utilization ratio during the checksum execution (\cref{table:checksum_eval}). As a reference for this ratio, we use the \emph{peak GPU performance}, which assumes that all SMs of the GPU are filled with the maximum number of active warps possible (64 for A100) and that the compute pipelines are filled optimally (i.e., one instruction being executed per clock cycle). Note that the number of warps, that are executed concurrently per clock cycle (4 for A100) is limited by the amount of compute units on the SM.

We compare our reference implementation from the first experiment (in SASS) with the same code written in PTX (virtual assembly), that has been processed using the \nvidia PTXAS assembler with the highest possible level of optimizations enabled. In comparison, optimized version of the checksum function that we generated using our instruction generation framework is around $\mathord{\sim}230\%$ faster than an implementation in PTX.

The checksum function in experiment 3 and 4 contains self-modifying code. This requires triggering cache eviction of the instruction cache such that the modified instruction gets updated. To trigger the cache eviction for the L2 instruction cache (\SI{128}{\kilo \byte}), the checksum loop is required to be larger than the cache size. As a consequence, we use 8342 \SI{16}{\byte} instructions in the checksum loop. With this cache eviction strategy, our implementation is able to achieve 75\% of the maximum utilization. Upon closer inspection with a GPU profiler, we find that that 99\% of all pipeline stalls, that happen during the execution of the checksum function, are caused by the fact that no instructions are available in the instruction cache to be executed. On average, each warp of this kernel spends 14.1 cycles being stalled due to not having the next instruction fetched yet. 
In comparison, reducing the size of the checksum loop to \SI{6.7}{\kilo \byte} (as in Experiment 1), we achieve a utilization of 99\% (without triggering cache eviction). This means that the hardware is unable load the updated (modified) instructions in-time for execution without causing any pipeline stalls. By comparing the VF's performance in experiment 1 and 3, we can conclude that a higher utilization can be achieved in case other cache eviction strategies become available to user code (e.g., specialized instruction).

In addition to the previous experiment, we modified the checksum function by adding an ``inner'' loop to the main loop of the checksum function calculation (Experiment 4). This effectively hides the performance loss due to cache misses in the instruction cache and achieves 100\% of the GPU peak performance. However, the time required to verify the code outside of the nested loop drastically increases and is thus considered too long to be practical.

\subsection{Attack Robustness}
To evaluate the robustness of our VF implementation with regards to attacks, we estimate the number of instructions that can be injected by an adversary without causing a noticeable time overhead. For this purpose, we measure the performance of the checksum function for 100,000 iterations and record the the standard deviation $\sigma$ of the total execution time based on 100 runs. 
We assume that the results of this experiment series are normally distributed and set the threshold value to detect adversarial tampering to be at $2.5 \cdot \sigma$ from the mean. The probability of a false positive is about $0.5\%$, in which case the verification process is restarted. 

To evaluate the robustness of this approach, we insert one additional NOP instruction in Experiment 2 (adversarial NOP) and report the minimum run time $T_{min}$ (averaged over 100 runs). Assuming a detection threshold of $T_{avg} + 2.5 \sigma$, we can conclude that $T_{avg} + 2.5 \sigma < T_{min}$ and thus it is impossible to insert one or more instruction without detectable overhead.

\subsection{Memory Region Inclusion Probability}
To evaluate how resilient our approach is regarding minor modifications in memory region containing the VF code (e.g., bit flips), we estimate the probability that a particular location is never included into the checksum result. We assume that memory accesses are distributed uniformly. Each block contains a single random memory access that loads an aligned 32-bit integer. For 100,000 iterations and a total checksum size of 524288 integers, the probability that a memory location is never included in the checksum result is negligible:
\[(1-1/524288)^{100000} = 0.082\]
%which is considered negligible.

\subsection{Execution of a User Kernel}
\name runs the original user kernel after verifying the code integrity, which preserves the original performance of the kernel. We evaluated the performance impact of running a user kernel using \name by implementing a simple benchmark based on matrix multiplication. Once the checksum computation is complete, the user kernel gets launched and is executed without modification. The baseline performance is the running time of the kernel, running without any verification.

We verified that the performance impact does not vary between small and large applications by running two different configurations of the benchmark, using small ($320x320$) and large ($6400x6400$) inputs. In our experiments, the performance impact of executing a user kernel in \name was negligible, apart from the expected time overhead of running the checksum function beforehand. \Cref{tab:user_kernel_perf} reports the measured number of clock cycles for each of the user kernel configurations, along with the baseline execution time and the verification overhead of the checksum (repeated once).

\begin{table}[h!]
	\small
	\centering
	\begin{tabular}{@{}lrrr@{}}\toprule
		Matrix size & Base & Verif. & SAGE \\ \midrule 
		$320 \times 320$ & $50576$ & $768 \times 10^6$ & $50499$ \\
		$6400 \times 6400$ & $215 \times 10^6$ & $757 \times 10^6$ & $215 \times 10^6$ \\ \bottomrule
	\end{tabular}
	\caption{Execution time (in clock cycles) of the user kernel and verification, compared to baseline time.}
	\label{tab:user_kernel_perf}
\end{table}

\subsection{Limitations of the Prototype}
The use of self-modifying code requires triggering cache eviction of the instruction cache such that the modified instruction gets updated. With this cache eviction strategy, our implementation is able to achieve 75\% of the maximum utilization. This is due to the GPU hardware not being able to load the required instructions in-time for processing after the L2 cache eviction. If other cache eviction strategies become available to user code, higher utilization can be achieved.
% For example, the \texttt{discard} keyword in PTX discards the L2 data and instruction cache, but the hardware still does not understand that it should reload the instructions.
Unfortunately, triggering cache eviction using a large checksum loop limits the time difference caused by an adversary inserting instructions into the checksum loop. 
%To reliably detect the insertion of a single instruction would require executing the checksum loop at least 1110 times resulting in an execution time of \SI{159.63}{\second}. Although this might only be practical for long running compute tasks.
We believe that GPU vendors with in-depth knowledge of GPU architecture would be able to reduce the checksum loop size and still use self-modification.
%, which would result in a lower execution time. 
Furthermore, the evaluation of our design demonstrates its technical feasibility.

\section{Security Analysis}
In the following, we systematically analyze potential attacks given our threat model (see~\cref{sec:threat-model}).

\paragraph{Pre-computation}
The result of the checksum function depends on an unpredictable challenge issued by the verifier enclave. This prevents pre-computation attacks where the checksum value or part of the checksum (e.g., intermediate values) are pre-computed to later run code other than the VF.

\paragraph{Computation optimizations}
The checksum function implementation must be time-optimal as algorithmic optimization would allow the adversary to find
computationally faster or more efficient way of computing the checksum value (see~\cref{sec:time-opt-requirements} for details). Given that our optimized version of the checksum computation achieves the GPU's peak performance, the computation steps cannot be optimized further.

\paragraph{Attacks on the host system}
The host system is untrusted (except for the verifier enclave)
and the adversary is assumed to have administrative control over the system.
This enables the adversary to eavesdrop, intercept, modify, or delay challenges or checksum
results being transmitted between the verifier and the device (e.g., on the PCI
bus). Given that the communication channel during the checksum computation is
unauthenticated, the adversary could also inject challenges or checksum results.
Modifications to the challenge would lead to a different checksum result.
By injecting challenges the adversary could treat the VF as an oracle;
however, given the unpredictable challenge generation, the probability of the verifier
reusing the same challenge value is negligible.

\paragraph{Attacks on the device / Resource takeover}
Before running the verification function, the device is considered untrusted. An adversary could be present on the device and interfere with the execution of the VF (e.g., by replacing or reordering instructions). This is prevented by the self-verification property and the strongly-ordered design of the checksum function.
A strongly-ordered function requires the adversary to perform the same operations on the same data in the same sequence as the original function to obtain the correct result. Otherwise, the output differs with high probability if operations that have dependencies among them are evaluated in a different order.

The adversary could also run computations on the device in parallel to the checksum computation. Our design uses all available SMs
simultaneously and maximizes thread and register usage. Thus, if an adversary would run a computation, the checksum computation would be deferred resulting in a considerable time overhead.
However, the execution of a user kernel might not require all available GPU resources (e.g., during data transfer to the GPU) and would allow the adversary to take over these available resources. To thwart these kind of attacks, the checksum function releases only the resources required for the execution of the user kernel using dynamic parallelism, and keeps the remaining resources (e.g., idle threads).
Furthermore, the use of cooperative full-kernel synchronization inside the user kernel guarantees that not other kernels are running, as the execution of two kernels would result in a deadlock.

\paragraph{Data substitution attacks}
\label{sec:data-sub-attack}
The adversary can try to modify memory locations of the VF while
keeping a correct copy of the modified values. During the execution of the
checksum function, when one of the modified memory locations is read, the adversary
redirects the read operations to the correct copy. In earlier work, this attack
is referred to as data substitution attack~\cite{seshadri2005pioneer,
seshadri2004swatt}. To maximize the time overhead introduced by this attack, the
data locations accessed by the checksum computation are
determined in a pseudo-random manner. This forces the attacker to monitor and
potentially redirect every memory read conducted by the checksum function.

\begin{figure}[tbh]
% \vspace{-0.1cm}
\centering
\includegraphics[width=0.9\linewidth]{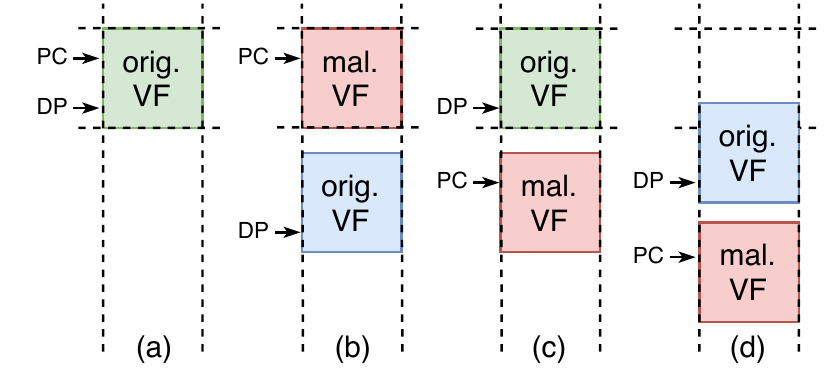}
% \vspace{-0.3cm}
\caption{Memory copy attack variants.}
\label{fig:mem_cpy_attack}
% \vspace{-0.2cm}
\end{figure}

\paragraph{Memory copy attacks}
\label{sec:mem-cpy-attack}
Seshadri~\etal~\cite{seshadri2005pioneer} specify memory copy attacks that can
be conducted by the adversary in the following three different ways as
illustrated in~\cref{fig:mem_cpy_attack}:
\begin{enumerate}[nolistsep]
%   \item[(a)] shows the unmodified VF with a correct program counter and data pointer.
  \item[(b)] the adversary replaces the checksum function
  with an
altered checksum function and executes it, but computes the checksum over a
correct copy of the checksum function elsewhere in memory. Thus, the program
counter is correct, but the data pointer points to the original copy of the
checksum function in a different memory location.
  \item[(c)] the adversary uses the correct checksum function code in the original
memory location to compute the checksum value, but executes a modified checksum
function elsewhere in memory. Thus, the data pointer points to the original
checksum function, but the program counter will be different.
  \item[(d)] the adversary places both the original checksum function code and its altered
version elsewhere from the memory locations where the correct checksum code
originally resided. Thus, both the program counter and the data pointer will be
different compared to an execution of the original checksum function.
\end{enumerate}

To prevent memory copy attacks, both the program counter and the data pointer need to be included in the computation of the checksum. The DP is included in each step of the computation, whereas the PC is indirectly included using self-modifying code.
In addition to these specified attacks, the attacker could also move the entire checksum function to a different location in memory using a deep memory copy. This would modify the position of the checksum function in the memory, but not its functionality. Thus, this is not considered a memory copy attack.

\paragraph{Proxy attacks}
We refer to proxy attacks as attacks where the adversary eavesdrops on the communication and obtains the challenge sent to the device, sends it to a proxy, computes the checksum function there and returns the result to the verifier.
We distinguish between the following cases:
\begin{enumerate}
    \item \emph{other GPU on the same host}: in case of a multi-GPU environment, we suggest to establish (and maintain) in sequence starting from the most powerful GPU to the least powerful one.
    \item \emph{other GPU on a different host}: By involving a remote entity, the measured execution time
will increase by the network latency for both sending the challenge and receiving the response. Tuning the number of checksum iterations to make the detection threshold smaller than the network latency, prevents using a more powerful GPU in a remote location. Furthermore, given that the verifier enclave frequently sends challenges and measures the execution time, this effect is amplified and thus would result in a noticeable difference with high probability~\cite{LiMcCunePerrigViper2011}.
\end{enumerate}

\paragraph{TOCTOU attacks / Execution environment takeover}
Time-of-check to time-of-use (TOCTOU) attacks~\cite{de2021toctou} are caused by a race condition involving the checking of the state of a part of a system and the use of the results of that check. In~\name{}, these attacks are considered because the checksum computation happens prior to the execution of the user kernel. In the particular, the adversary has two points where it could take over the execution environment set up by the VF: 1) before the launch of the user kernel, and 2) after the execution of the user kernel has completed. In the former case, the execution of the user kernel gets scheduled by the scheduler on the GPU. In case another kernel that is controlled by the adversary is present, this kernel could be scheduled instead of the user kernel. This is prevented by inlining the user kernel into the VF such that the epilog of the VF can directly call the user kernel using a function call. In the latter case, the execution of the user kernel has finished and thus the user is indifferent whether the dynamic root-of-trust has been compromised. If the user wants to execute another kernel, the dynamic RoT needs to be re-established.

\paragraph{Physical attacks}
Our threat model considers physical attacks on the server. Consequently, the adversary could get physical access to the host and also the GPU memory, perform a memory dump, and analyze it to extract potential secrets. Unlike snooping and tampering off-package memory, local memory in a GPU package is considered to be much harder~\cite{conductor-security}. To increase the attack complexity of such physical attacks, the symmetric key established using SAKE is located in shared memory.
Physical attacks on the host system are also possible and could include snooping on the PCIe bus or the host main memory. Since the host system is assumed to be untrusted (except for the verifier enclave and its related components), physical attacks on the host would allow similar attacks as in software.

\subsection{Formal Verification of Modified SAKE}
To show that our modified SAKE protocol securely establishes a key between the verifier and the GPU, we have formally modeled the key establishment protocol and verified its security properties using the Tamarin prover~\cite{tamarin} under the assumption that the computed checksum provides a short-lived secret.
To model this property in Tamarin, we use a single-use authentic channel over which we will send $w_2, \text{MAC}_c(w_2)$.
We show that the established symmetric key remains secret and is unique, a weak agreement exists between the verifier and the device, and recent aliveness for each run of the protocol~\cite{tamarin-manual-property}.
% Weak agreement denotes that whenever an agent $v$ in the role of the verifier $V$ has completed a run of the key exchange, then another agent $d$ in the role of the device $D$ has been running the key change protocol, apparently with $v$. Recent aliveness describes the property that for each run of the key exchange the device $D$ has recently been alive.
The proofs generated using the Tamarin prover can be found here \url{https://github.com/spcl/sage/tree/main/proofs}.

\section{Related Work}
\label{sec:related}

% \subsection{Trusted Execution}
% Mainstream processor vendors have implemented trusted execution environments
% (TEEs), such as Intel Software Guard Extensions (SGX)~\cite{costan2016intel},
% ARM TrustZone~\cite{arm-trustzone}, Keystone~\cite{lee2020keystone}, and
% Sanctum~\cite{costan2016sanctum}, in most of their state-of-art chips. These
% TEEs isolate a secure world from the regular execution environment, such that
% protected data can be processed in the secure world. However, none of these TEEs
% are designed to support trusted execution on GPUs.

\paragraph{Trusted execution on GPUs}
To support trusted execution on GPUs, the following approaches were proposed.
Graviton~\cite{volos2018graviton} specifies an architecture for supporting
trusted execution environments on GPUs by changing the GPU's command processor
to perform remote attestation based on device specific keys and ensure isolation
between multiple processes running on the GPU. This is achieved by
utilizing a set of keys where the root key gets baked into device upon its
creation. The latter requires modification to the GPU hardware by modifying the
GPU's internal command processor to impose a strict ownership discipline.

HIX~\cite{jang2019heterogeneous} proposes a heterogeneous isolated execution
environment. HIX does not require modifications to the GPU architecture to offer
an isolated execution environment, but instead physically modifies the I/O
interconnect between the CPU and GPU and refactors the GPU device driver to work
from within a TEE on the host. The TEE can then allocate trusted enclaves on the
GPU.

HETEE~\cite{zhu2019enabling} is based on a standalone computing system to
dynamically allocate accelerators (such as GPUs or FPGAs) for either secure
computing, or available to the host OS using PCIe switches. The security
controller (and its software) is assumed to be trusted and interacts with the
management CPU to control PCIe switching. HETEE attempts to provide isolation by
selectively making accelerators available to specific applications by controlling
communication to the accelerator through the security controller.

Telekine~\cite{hunt2020telekine} illustrates side-channel attacks against TEE on GPUs based on observing the timing of GPU kernel execution. It then introduces a GPU stream abstraction that ensures execution and interaction through untrusted components are independent of any secret data. Telekine requires a GPU TEE to be deployed.

%BorderControl~\cite{olson2015border}\ben{TODO.}

% \paragraph{Privacy-preserving machine learning}
Machine learning represents a major use case for using GPUs as accelerators and
can require privacy-preserving approaches for sensitive data.
Slalom~\cite{tramer2018slalom} uses a combination of a trusted enclave and
untrusted GPU. The system decomposes the machine learning into two parts, where
the control flow part runs inside the trusted enclave and operations that are
not privacy sensitive (such as convolutions based on matrix multiplications) are
offloaded to the GPU. Unfortunately, the split results in a decrease of
training and inference accuracy. 
%\adrian{I thought there is also a significant slowdown ...}\ben{they present the result in comparison to execution on the CPU. Compared to an execution on the GPU, they would probably have a slowdown, but provide no numbers in their paper}

% Myelin~\cite{hynes2018efficient} computes entire ML workloads in SGX enclaves
% which cannot achieve the efficiency of using accelerators.

% SecureML~\cite{secureML} is based on a two-server model where data owners
% distribute their private data among two non-colluding servers for privacy. The
% servers train various models on the joint data using secure two-party
% computation.

%CryptGPU: Fast Privacy-Preserving Machine Learning on the GPU

% \subsection{Software-based Attestation}
\paragraph{Software-based attestation} %
% Apart from code attestation based on hardware (e.g.,
% TCG~\cite{sailer2004design}, SGX~\cite{costan2016intel}) also approaches purely
% based on software have been proposed.
SWATT~\cite{seshadri2004swatt} uses a verification function that is based on
pseudo-random memory traversal to a compute memory checksum. The verifier
measures the execution time and verifies the checksum. Malicious
code is required to verify each memory access to replace memory reads of
changed locations with expected content, resulting in detectable time
overhead. SWATT checks the entire memory of a system and its running time becomes
prohibitive on systems with large memories.

PIONEER~\cite{seshadri2005pioneer} verifies the integrity and guarantees the
execution of code using a checksum function that is closely tied to the
Pentium 4 architecture.
% The verification function contains three parts: a
% checksum code, a hash function and a send function.
The checksum function
computes a fingerprint of the verification function and sets up an untampered
execution environment. It is constructed such that
manipulations by the adversary will noticeably increase the computation time.
% The hash function is used to measure the integrity of the executable, whereas
% the send function is used to transfer the checksum and integrity measurement
% back to the verifier.

Kovah et al.~\cite{Kovah-oakland12} and Butterworth et al.~\cite{Butterworth-ccs13} extended the checksum computation to work on a Microsoft Windows system (CPU only), enabling a remote verifier to attest to a running system in a corporate environment. 
% Their system only verifies code running on the CPU, not on any peripheral.

Shaneck et al.~\cite{shaneck2005remote} describe a software-based approach to remotely attest the static memory contents of sensors without requiring any additional hardware on the sensors nor precise measurements of execution timing. They use self-modifying code that generates memory read and jump instructions during the execution of their code.

% Program-Integrity Verification (PIV)~\cite{} verifies the integrity of code residing in a sensor using a randomized hash function whenever the sensor joins a network.

% SWATT~\cite{seshadri2004swatt} uses a verification function that is based on
% pseudo-random memory traversal to a compute memory checksum. The verifier
% measures the execution time and verifies the checksum. This leads to malicious
% code being required to verify each memory access to replace memory reads of
% changed memory locations with expected content, resulting in detectable time
% overhead. SWATT checks the entire memory of a system and its running time becomes
% prohibitive on systems with large memories.

% ReDABLS~\cite{zhao2013redabls} revisits the notion of device attestation with
% bounded leakage of secrets (DABLS) to overcome practical challenges by
% introducing much lower overhead rates, particularly when acceptable
% probabilistic upper bounds are found for an adversary’s success in attacking it.

Gligor and Woo~\cite{gligor2019establishing} proposed a system that allows to
provably establish a root of trust and provide secure initial states for all
software unconditionally.
% without requiring any pre-established secret, trusted hardware modules or special instructions. 
The authors design a family of $k$-independent (almost)
universal hash functions based on polynomials and use Horner's rule to show
time- and memory-optimal evaluation of polynomials.
% To hash the memory contents, each word in memory gets xor'ed to a coefficients of the polynomial.
Their proofs only hold under the unit time assumption in their concrete Word Random Access Machine (cWRAM) model. In addition, the proposed polynomials get very large for large number of
words (computation time in the order of minutes).
%the device is assumed to be directly connected to the bus (not possible without HW modifications)

\section{Conclusion}
\label{sec:conclusion}

The prospect of software-only trust root establishment and secure code
execution on GPUs offers exciting opportunities: execution of
sensitive GPU code that should not be leaked to the GPU operator (code
secrecy), correct execution of GPU code in an adversarial environment
(code and execution integrity), preserving data correctness and
confidentiality in the presence of malicious code on the system (data
secrecy and integrity). \name represents a first step for achieving
these properties on the \nvidia Ampere architecture, under the
circumstances that the architectural details about the Ampere architecture are closed-source. Since architectural knowledge for designing the verification function (VF) is key, our software-based approach to provide secure code execution on GPU paves the way forward for GPU vendors: they are naturally in a position to align the design of the VF to their architectural knowledge and lead the standardization process for trust establishment on GPUs.

Remaining open challenges include the design of software-based secure
execution on alternative platforms, improving the execution speed of
the verification function, and extend the execution model to support
libraries that use a hybrid CPU+GPU compute model (e.g., TensorFlow~\cite{tensorflow}). 
Ultimately, an interesting future research question to
answer is the interplay between hardware- and software-based
approaches for trusted execution to achieve the strongest possible
security properties for GPU-based execution.

\clearpage
\bibliographystyle{IEEEtran}
\bibliography{references.bib}

\end{document}